\documentclass[a4paper,11pt]{article}
\usepackage{pos}
\usepackage{subcaption}
\usepackage{wrapfig}

\title{Observation of microquasars high-energy emission with INTEGRAL}

\author*[a]{T. Bouchet}
\author[a]{J. Rodriguez}
\author[b]{F. Cangemi}
\author[a]{P. Laurent}

\affiliation[a]{DAP-AIM, IRFU, CEA, Université Paris-cité, CEA Saclay, France}

\affiliation[b]{APC, Université Paris-cité, France}

\emailAdd{tristan.bouchet@cea.fr}

\abstract{Microquasars are Black Hole X-ray binaries (BHXB) which can eject material in the form of a bipolar jet, similarly to quasars, but at much smaller scales. Their high-energy emission comes from an accretion disk ($\sim$1 keV) and from a hot \textquoteleft corona' near the black hole that up-scatters photons from the disk in the hard X-ray domain (1--100 keV). A high-energy component above 150 keV has been detected in bright sources and its precise origin is still unknown: it could come either from Compton scattering of disk photons on coronal relativistic non-thermal electrons (a.k.a hybrid Comptonization), or from the synchrotron emission from the very base of the compact jet.
The measurement of polarization above 150 keV can provide valuable insights into the processes at play as we expect higher polarization fraction due to synchrotron emission from the jets (up to 70\% with a very ordered magnetic field)}

\FullConference{High Energy Phenomena in Relativistic Outflows VIII (HEPROVIII)\\
 23-26 October, 2023\\
Paris, France\\}

\begin{document}
\maketitle

\section{The INTEGRAL satellite}

INTEGRAL (INTernational Gamma-ray Astrophysical Laboratory) has been observing the soft gamma-ray sky for more than 20 years, thanks to its different instruments: the IBIS (15 keV -- 10 MeV) spectro-imager with its high spatial resolution (12’), the SPI spectrometer (20 keV –- 8 MeV), capable of probing the soft gamma-ray spectra in great detail, and the JEM-X (3 -- 30 keV) X-ray monitor, which can help constraining the soft X-ray emission.\\

\section{Compton mode and polarization}
\begin{wrapfigure}{R}{0.5\textwidth}
\begin{center}
    \includegraphics[scale=0.35]{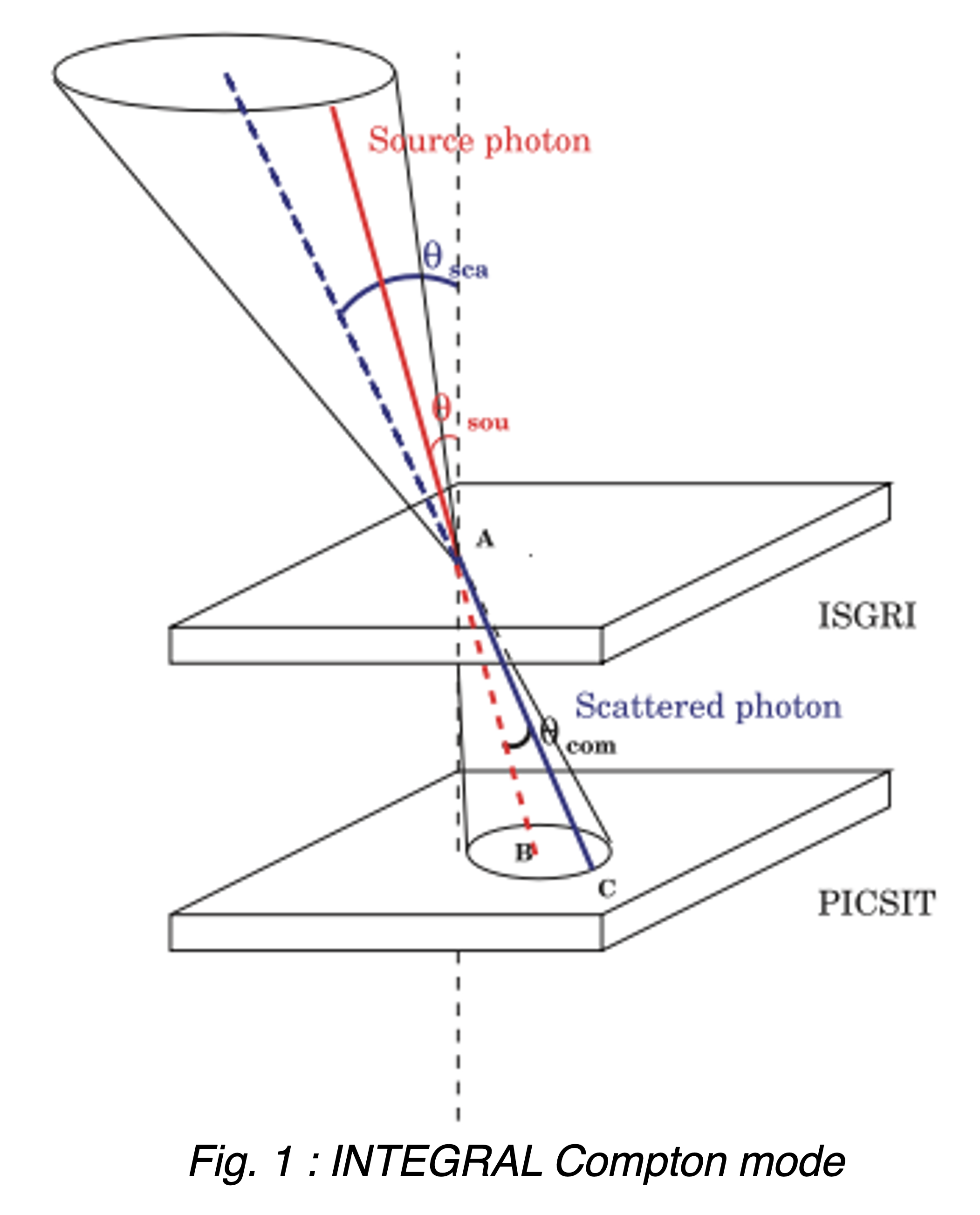}
    
\end{center}
    \caption{Sketch of IBIS Compton mode on board INTEGRAL (adapted from \cite{Forot_2007})}
    \label{fig:forot_sketch}
\end{wrapfigure}

It is possible to combine the two detector layers of the IBIS telescope (ISGRI and PICsIT, see Fig.\ref{fig:forot_sketch}) and use IBIS as a Compton telescope. Source fluxes are reconstructed through Compton scattering of photons on the first detector (ISGRI), then absorbed by the second detector (PICsIT) \cite{Forot_2007}.\\

We deduce the total energy of the photon from the energies deposited in each detector. The source position is determined with coded mask techniques using the Compton/ISGRI image. This technique allows us to look at photons between 200 and 2000 keV. This observation mode has been calibrated using a Monte-Carlo GEANT3 simulation. The rate of spurious events (two independent photons detected simultaneously by both detectors) has also been carefully modeled and taken into account.\\

\begin{figure}
    \centering
    \includegraphics[scale=0.5]{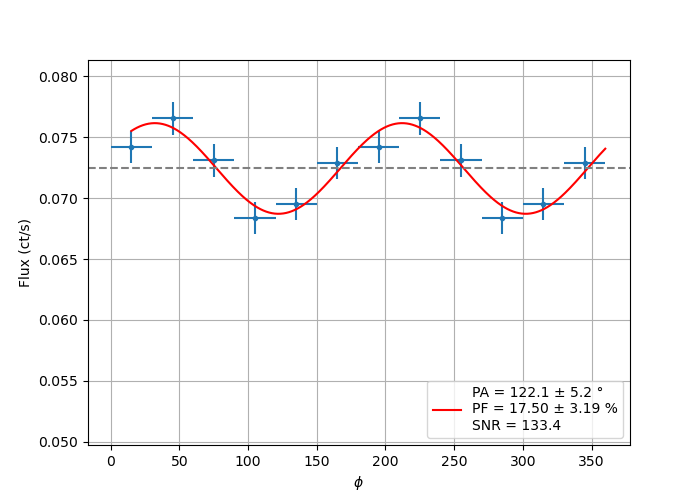}
    \caption{Modulation of the Crab nebula flux as a function of the azimuthal angle, also known as polarigram, in the 250 -- 450 keV band.}
    \label{fig:crab_polarigram_250_450}
\end{figure}

For Compton events, the spatial distribution of the photon on PICsIT is affected by the source polarization.
The azimuthal angle of the scattered photon ($\phi$) will be modulated as:

\begin{equation} \label{eq:1}
N(\phi)=C\ (1+a_0\ cos{(2(\phi-\phi_0))})
\end{equation}
with C the total count-rate and $a_0$ the amplitude of the modulation ($a_0$ = 0 for an unpolarized source). From this asymmetry in angle distribution, we can deduce the Polarization Angle: $PA = \phi_0 - \pi/2$, and the Polarization Fraction: $PF= a_0 /a_{100}$  with $a_{100}$ the amplitude for a fully polarized source (computed through simulations). This method has been applied for several sources before \cite{Forot_2008, Laurent_2011, Cangemi_2023} and gave compatible results with other observatories at different wavelengths.\\
We checked again the consistency of our method using the Crab Nebula, a bright and relatively stable source, making it a \textquoteleft standard candle' for X-ray astronomers. We accumulated 9 Ms of Crab observations from 2005 to 2019 to test the validity of our method. We found PA=(122~±~5)° [modulo 180°] and PF=(17~±~3)\% in the 250 -- 450 keV band (see Fig.\ref{fig:crab_polarigram_250_450}). This is in good agreement with a previous measurement using an independent method with the INTEGRAL/SPI instrument, which found a PA of (120~±~6)° and a PF of (24~±~4)\% in the 130--436 keV band, over the same time period \cite{Jourdain_2019}.

\section{The new source Swift J1727.8-1613}
Microquasars typically enter in outburst for a few months and then go back to quiescence for many years.
A new source named Swift J1727.8-1613 was discovered in August 2023 with a particularly high flux ($4.2~ 10^{-8}$ erg/cm$^2$/s in the 30 -- 100 keV energy range) and was quickly identified as a BH X-ray binary \cite{s1727_maxi_atel}.\\

\subsection{Spectrum}
\begin{figure}
    \centering
    \includegraphics[scale=0.5]{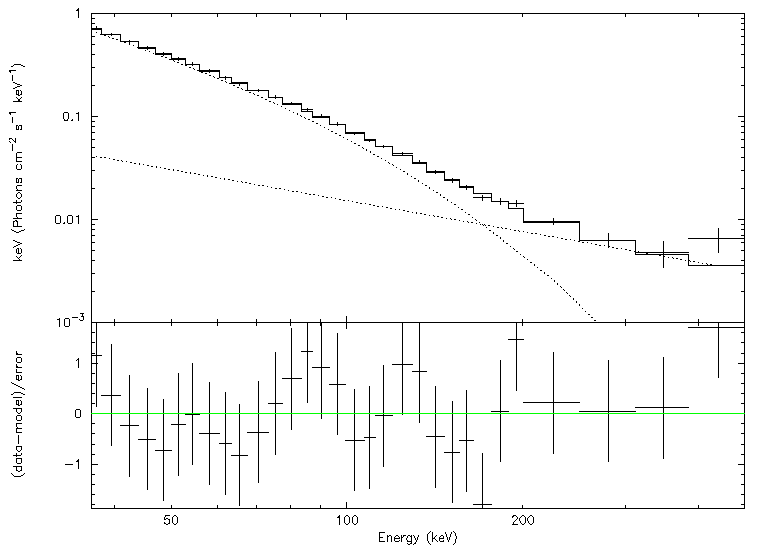}
    \caption{INTEGRAL/ISGRI spectrum of the microquasar Swift J1727.8-1613 in the HIMS.}
    \label{fig:isgri_him}
\end{figure}

We looked to the spectra of this source with the ISGRI detector and accumulated data during its Hard-Intermediate State (HIMS).
As shown in Fig.\ref{fig:isgri_him}, we found the dominant component below 100 keV to be well fitted by a cut-off power law, with a cut-off energy around 60 keV, which is a good representation of the Comptonized emission from a hot thermalized corona. The spectral shape of this emission was stable throughout the source HIMS observations. A hard-tail is clearly detected until at least 500 keV, which is well fitted with a power law with a photon index $\Gamma\sim 1.8$ \footnote{$dN(E)\propto E^{-\Gamma} dE$, N the number of detected photons/cm$^2$/s/keV.}.\\

\subsection{Polarization}

\begin{figure}
    \centering
    \includegraphics[scale=0.6]{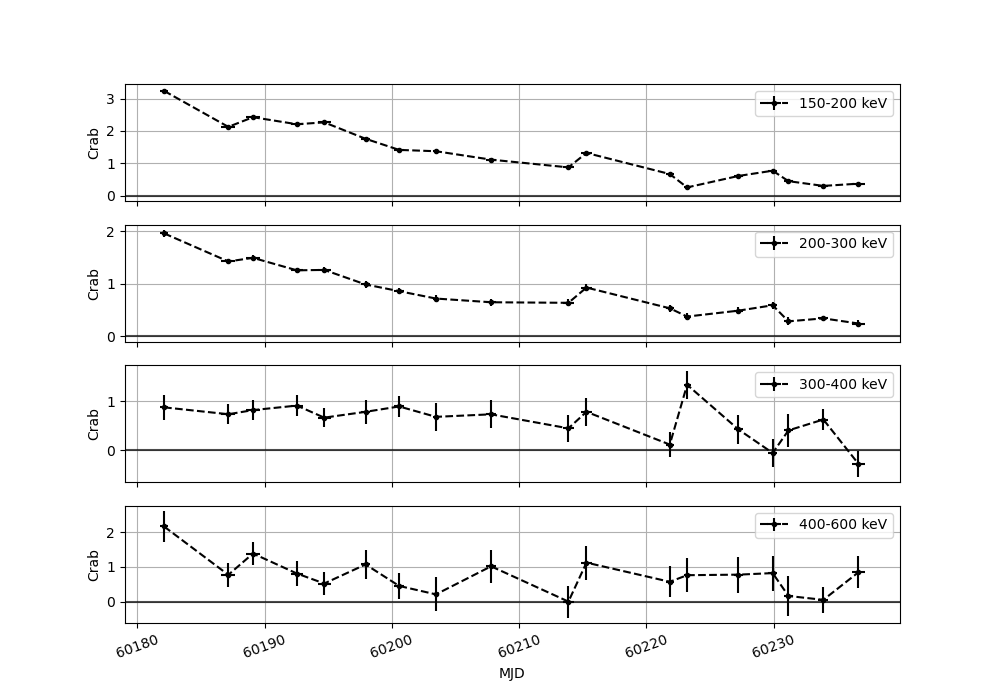}
    \caption{IBIS/Compton mode flux evolution of Swift J1727.8-1613 in Crab units, in different energy bands.}
    \label{fig:S1727_compton_lc}
\end{figure}

\begin{figure}
    \centering
    \includegraphics[scale=0.4]{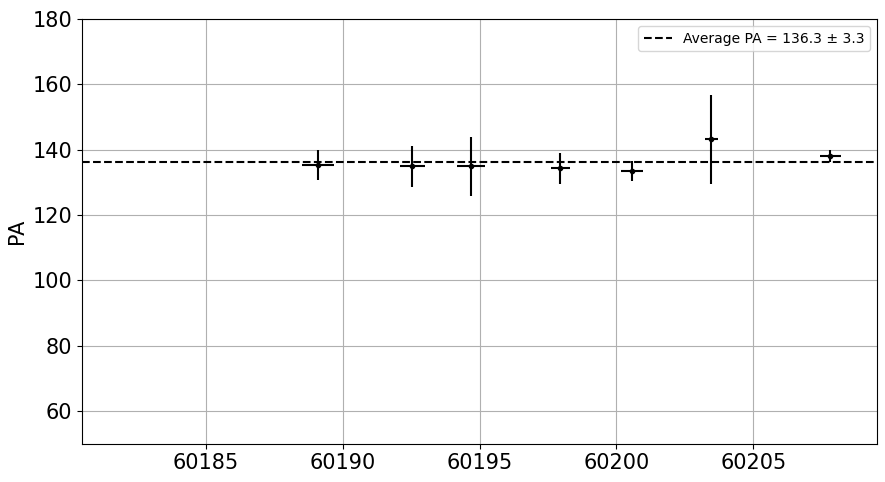}
    \caption{Evolution of Polarization Angle of Swift J1727.8-1613 with time (MJD).}
    \label{fig:hims_pola_evolution_200_260}
\end{figure}

Thanks to the source high brightness we were able to monitor the high-energy flux obtained with the IBIS/Compton mode (see Fig.\ref{fig:S1727_compton_lc}), as well as its polarization angle in the soft gamma-ray band.
The 200 -- 400 keV hard power-law flux decreases progressively throughout the outburst in all bands, following the usual decrease in hardness observed during HIMS \cite{Belloni_2010}. In the 200 -- 260 keV band, we can follow the polarization on revolution basis ($\sim$ 2.5 days): as shown in Fig.\ref{fig:hims_pola_evolution_200_260}, the polarization angle stays around a stable value of PA = (136~±~3)° [modulo 180°].\\

\begin{figure}[h]
    \centering
    \subfloat[200 -- 300 keV band]{
    {\includegraphics[scale=0.45]{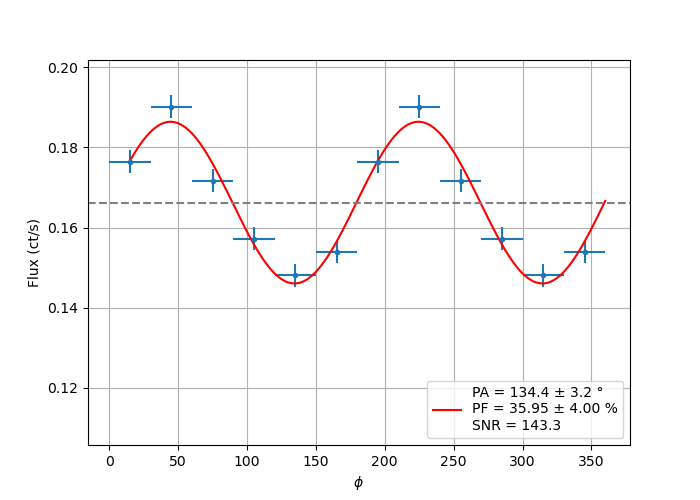} }}%
    \subfloat[300 -- 400 keV band]{
    {\includegraphics[scale=0.45]{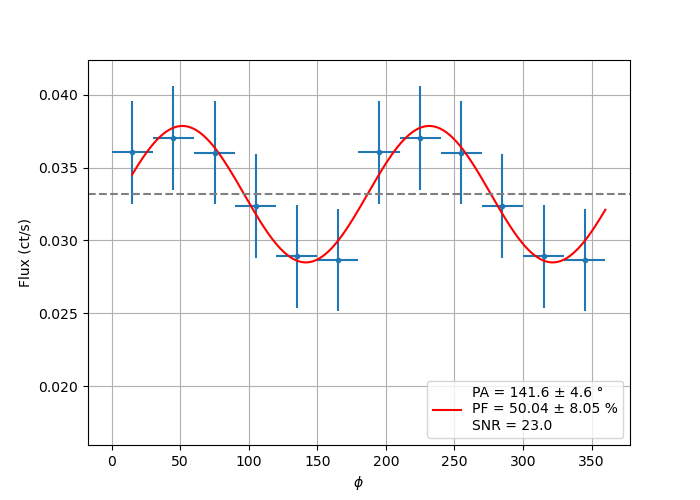} }}%
    \hspace{0mm}
    \caption{Polarigrams of Swift J1727.8-1613 in different energy bands in the HIMS.}%
    
\label{fig:S1727_polarigram}

\end{figure}

Similar polarization angles were found in higher energy bands; we measured a PF of (35~±~5)\% in the 200 -- 300 keV band, and (50~±~8)\% in the 300 -- 400 keV band (see polarigrams in Fig.\ref{fig:S1727_polarigram}). As seen in the continuum fitting, this energy domain is already dominated by the hard-tail, which would hint towards a compact jet origin for this $>200$ keV component.

\section{Conclusion}

For the very first time, we were able to follow the polarization in the soft-gamma ray band throughout the Hard Intermediate State of a microquasar. The polarization angle remained stable at values between 134° and 141° in the 200--400 keV band, while the polarization fraction reaches up to 50\%, which is compatible with optically thin synchrotron emission. 

\acknowledgments
The authors acknowledge partial funding from the French Space Agency (CNES). We thank Carlo Ferrigno for his help on the calibration of the IBIS/ISGRI detector. Based on observations with INTEGRAL, an ESA project with instruments and science data center funded by ESA member states (especially the PI countries: Denmark, France, Germany, Italy, Switzerland, Spain) and with the participation of Russia and the USA.

\bibliographystyle{JHEP}
\bibliography{biblio}

\end{document}